\documentclass[preprint]{jpsj3}
\usepackage{ulem}
\usepackage{color}
\usepackage{braket}
\usepackage{amsmath}
\definecolor{mygreen}{rgb}{0.0,0.7,0.0}
\newcommand{\txr}[1]{\textcolor{black}{#1}}

\title{ Resonant Inelastic X-Ray Scattering Spectra of Cuprate Superconductors Predicted by Model of Fractionalized Fermions}
\author{Masatoshi Imada}
\inst{Waseda Research Institute for Science and Engineering, Waseda University, Okubo, Shinjuku, Tokyo, 169-8555, Japan. \\
Toyota Physical and Chemical Research Institute, Yokomichi, Nagakute, Aichi, 480-1118, Japan.} 
\abst{We theoretically analyze the intensity of the resonant inelastic X-ray scattering (RIXS) based on Ansatz that the electron is fractionalized into two components. By fitting the angle resolved photoemission spectroscopy (ARPES) data of a cuprate high-$T_{\rm c}$ superconductor Bi$_2$Sr$_2$CaCu$_2$O$_{8+\delta}$  and the accompanied machine learning results obtained before by Yamaji {\it et al.} (arXiv:1903.08060), we determine the paprameters of the two-component fermion model which describes the electron fractionalization.  The RIXS spectra are predicted by using this formulation. We find that the intensity is enhanced in the superconducting phase relative to the normal or pseudogap phase. Since the enhancement is unusual, we propose this can be used as a critical test whether the unconventional electron fractionalization captures the essence of the cuprate superconductivity. This type of combined analyses using multiple independent spectroscopy data, with the help of theoretical and machine learning insights, regarded as {\it integrated spectroscopy} opens a new route to study difficult open issues in strongly correlated electron systems.  
}

\begin{document}
\maketitle

%
\section{Introduction}
Since the discovery of  copper oxide (cuprate) superconductors in 1986\cite{Bednorz1986},  
marvelous progress has been made toward the goal of understanding the mechanism of the superconductivity and the normal state properties exhibiting the pseudogap near the Mott insulating phase. Momentum and energy resolved spectroscopic experiments including the angle resolved photoemission spectroscopy (ARPES)\cite{Shen},  the quasiparticle interference (QPI)~\cite{Hoffman2002} at the scanning tunnel microscope (STM), and resonant inelastic X-ray spectroscopy (RIXS)~\cite{RIXS}, in addition to conventional probes such as optical measurements, neutron scattering are typical tools to elucidate the momentum/energy resolved dynamics of strongly correlated electrons. Despite their achievements, the mechanisms promoting the superconductivity and the pseudogap have not reached consensus.

On the other hand, {\it ab intio} framework of electronic structure calculations recently developed for strongly correlated electron systems~\cite{Imada_Miyake2010} has opened a route to understand high-temperature strong-coupling superconductivity in examples of the cuprate~\cite{Ohgoe2020} and iron-based superconductors~\cite{Misawa2014}, where the calculated phase diagrams have shown quantitative agreements with those of experiments without adjustable parameters after deriving the low-energy lattice Hamiltonians with both the kinetic and interaction strengths. 

One of the important insights gained from the {\it ab initio} studies for the origin of high-$T_{\rm c}$ superconductivity is the commonly identified strong effective attraction of electrons caused near the Mott insulator both for the cuprates and the iron-based superconductors (and in the iron-based case, proximity of additional high-spin states caused by the Hund's rule coupling), in other words, caused in correlation induced bad metals. This {\it attraction} is paradoxically caused by the original strong Coulomb {\it repulsion}  in the evolution at the release of carriers from incoherence.  This attraction causes severe competition toward  the real space charge inhomogeneity as well including charge order against superconductivity. 
It was proposed that the strong attraction and severe competition may also result in the underlying electronic bistability at double-well like energy structure as a function of the electron filling. 
This naturally accompanies resultant fractionalization of a bare electron into the conventional quasiparticles and dark fermions~\cite{ImadaSuzuki}.  Although the charge or spin may not necessarily have a fractional number in the present definition of fractionalized particles unlike the fractional quantum Hall effect, the fractionalization here implies that the bare electron splinters into constituents or splits into two or more degrees of freedom, which essentially behaves as independent fermi particles. The splintered two degrees of freedom may naturally represent the bistability as well.    

From a related but independent perspective, theoretical and numerical studies were conducted based on the cluster dynamical mean-field studies of the Hubbard model~\cite{Sakai2016a}. They revealed that the superconductivity in the Hubbard model,  which is adiabatically connected to the superconducting state found in the solution of {\it ab inito} Hamiltonians for a hole-doped cuprate compound, HgBa$_2$CuO$_4$~\cite{Ohgoe2020}, has a marvelous property: Each of the normal and anomalous self-energies has a prominent peak in the imaginary part  at the same energy in the $d$-wave superconducting state, while their peak contributions strictly cancel in the single-particle spectral function $A(k,\omega)$ so that the anomalous peak does not show up, consistently with the known ARPES measurement.    
Furthermore, the peak in the anomalous self-energy is identified as the real origin of the high-$T_{\rm c}$ superconductivity in the Kramers-Kronig analysis. 

This seemingly puzzling cancellation supports the idea that the underdoped region of the cuprates exhibits the fractionalization consistently with the above insight from the {\it ab initio} study of the effective Hamiltonians. This is because that if an electron fractionalizes into the conventional quasiparticle and a dark fermion, the resultant two-component fermion model (TCFM) remarkably shows exact cancellation described above, whereas the single-component fermions in the absence of the strong correlation effect do not show such a cancellation. 
The prominent peaks in the imaginary part of the normal and anomalous self-energies are understood from the pole of the dark fermion Green's function generating the self-energy pole of the quasiparticle Green's function (the imaginary part of which, namely the spectral function $A(k,\omega)={\rm Im} G(k,\omega)$ is measurable by ARPES) through the hybridization of the dark fermion and the visible quasiparticle.
In fact, the emergence of the pseudogap in the normal state is consistently understood from the remaining pole in the normal self-energy, which does not cancel with the anomalous self-energy part any more, because the anomalous part becomes vanishing in the normal phase. This pole in the pseudogap state has a structure similar to that proposed from the resonating valence bond picture for the pseudogap phase~\cite{YRZ}.

Furthermore, the machine learning study on the ARPES data in the superconducting states of Bi$_2$Sr$_2$Cu$_2$O$_{6+\delta}$ (Bi2201) and Bi$_2$Sr$_2$CaCu$_2$O$_{8+\delta}$ (Bi2212) has succeeded in inferring and extracting the normal and anomalous self-energies separately purely from the experimental data quite independently of the above model studies  and has shown consistency with the peak formation and the cancellation. This consistency again supports the fractionalization~\cite{Yamaji2021}. 

However, because this cancellation has a strong impact and poses a severe constraint on the mechanism of superconductivity, it is desired to gain further insights whether the fractionalization has consequences in other experimental probes especially in momentum-energy resolved spectroscopic measurements to deepen understanding from different angle of viewpoint.   
  
For decades, the strong correlations of electrons hamper the full  and unified understanding of the electron dynamics and resultant physical phenomena even with recent revolutionary improved each spectroscopic methods such as ARPES, QPI and RIXS. 
However, we may have hope to overcome the difficulty, if different and independent experimental tools are combined as {\it integrated spectroscopy}. The purpose of this paper is to initiate such an attempt of the integrated spectroscopy, by proposing an exemplified combination of the ARPES experimental data~\cite{kondo2011disentangling} together with results obtained by the machine learning analyses~\cite{Yamaji2021}, with the RIXS measurement using the simple TCFM based on the fractionalization of electrons. Especially in this paper, I focus on the RIXS spectra to gain insight into this enigmatic issue by utilizing the already analyzed ARPES data~\cite{Yamaji2021}. RIXS is a spectroscopic method of measuring the electron-hole excitation, which may also show a signal of the fractionalization as we discuss in this paper. 

So far we do not have sufficient and comprehensive data of RIXS measurement, which is rapidly developing with remarkably improved resolution and here I predict the unique qualitative behavior and dynamics of excitons in the RIXS particularly on the charge dynamics if the fractionalization is valid and the consequential TCFM essentially captures the exciton dynamics: I show that the RIXS intensity weight shows distinct enhancement in the superconducting than normal states if the fractionalization is evolved in the normal state.

In Sec.2, we introduce the TCFM. The parameters of the model are determined so as to reproduce the ARPES data for Bi2212~\cite{kondo2011disentangling} and its analysis by the machine learning~\cite{Yamaji2021}. In Sec.3, theoretical formulation for RIXS quantities in the experimental measurements is summarized.
In Sec.4, theoretical predictions on the RIXS spectra based on the TCFM derived in Sec.2 are presented.  
Sec.5 is devoted to discussions including them on the mechanism of the enhancement.
In Sec.6, I conclude the paper with future outlook.
 
%
\section{Two-Component Fermion Model (TCFM)}
%
The TCFM for the cuprate superconductors is represented by the Hamiltonian
\begin{eqnarray}
H&=&\sum_{k,\sigma}[ \epsilon_c (k)c_{k,\sigma}^{\dagger}c_{k,\sigma} +\epsilon_d (k)d_{k,\sigma}^{\dagger}d_{k,\sigma}  \nonumber 
\\
&+& \Lambda (k) (c_{k,\sigma}^{\dagger}d_{k,\sigma} +{\rm H.c.})
\nonumber 
\\
&+&(\Delta_c(k) c_{k,\sigma}^{\dagger}c_{-k,-\sigma}^{\dagger}+\Delta_d(k) d_{k,\sigma}^{\dagger}d_{-k,-\sigma}^{\dagger} + {\rm H.c})
]. \nonumber \\
\label{TCfermionAnomalous} 
\end{eqnarray}
 Here, the dark fermion represented by the annihilation operator $d$ with the dispersion $\epsilon_d(k)$ is hybridizing with the bare electron or quasiparticle ($c$ being its annihilation operator) with the dispersion $\epsilon_c(k)$ at the momentum $k$ in a form of a noninteracting Hamiltonian, where the hybridization is represented by $\Lambda(k)$ term. 
In this paper, we employ a square lattice with simple dispersion for $\epsilon_c (k)$ and $\epsilon_d (k)$ as
\begin{eqnarray}
\epsilon_c(k)=-(2t_{c1}(\cos k_x+\cos k_y)+4t_{c2}\cos k_x\cos k_y)+\mu_c, \nonumber \\
\epsilon_d(k)=-(2t_{d1}(\cos k_x+\cos k_y)+4t_{d2}\cos k_x\cos k_y)+\mu_d,
\label{dispersion} 
\end{eqnarray} 
where $t_{c1}$ ($t_{d1}$) and $t_{c2}$ ($t_{d2}$) represent the nearest-neighbor and next-nearest-neighbor hoppings of $c$ ($d$) fermion, respectively.
The terms proportional to $\Delta_c(k)$ and $\Delta_d(k)$ represent the anomalous part in the mean-field approximation emerging in the superconducting state. Here we assume the simple $d$-wave gap 
\begin{eqnarray}
\Delta_c(k)=\frac{\Delta_{c0}}{2}(\cos k_x-\cos k_y), \nonumber \\
\Delta_d(k)=\frac{\Delta_{d0}}{2}(\cos k_x-\cos k_y),
\label{SCterm} 
\end{eqnarray} 
with temperature dependent constants $\Delta_{c0}$ and $\Delta_{d0}$. 
Note that the pseudogap is the consequence of the fractionalization of an electron into the quasiparticle $c$ and the dark fermion $d$ as we describe below and only the low-energy part  near the Fermi level has mapping between the $c$ and $d$ fermions and the original electron by exhausting the degrees of freedom in the Hilbert space~\cite{ImadaSuzuki}. The higher energy part far from the Fermi level contained in the upper and lower Hubbard bands is not taken into account in this simple TCFM.  

 In the normal state ($\Delta_c(k)=\Delta_d(k)=0$), the direct hybridization gap at each momentum in the normal state is given by
\begin{eqnarray}
\Delta_{\rm HG}(k) =\sqrt{(\epsilon_c(k)-\epsilon_d(k))^2+4\Lambda (k)^2}.
\label{HGap} 
\end{eqnarray} 
representing the pseudogap. After introducing a phenomenological damping $\gamma(\omega)$, Green's function in Nambu representation is obtained from the 4x4 matrix inversion as
\begin{eqnarray}
G(k,\omega)=(\omega-H-iI\gamma(k,\omega))^{-1},
\label{TCMNGkw} 
\end{eqnarray}
 with $H$ given by Eq.(\ref{TCfermionAnomalous}) and $I$ being the 4 by 4 identity matrix. Then  normal Green's function for $c$ is given from the (1,1) component in the form 
\begin{eqnarray}
G_c(k,\omega)=\frac{1}{\omega-\epsilon_c(k)-\Sigma (k,\omega)},
\label{TCMNGkw} 
\end{eqnarray}
with the self-energy
\begin{eqnarray}
\Sigma(k,\omega)=\frac{\Lambda (k)^2}{\omega-\epsilon_d(k)},
\label{TCMNSkw} 
\end{eqnarray}
where the pseudogap is represented by the suppressed density of sates caused by the pole of the self-energy at $\omega=\epsilon_d(k)$.

In the superconducting state, the normal part of the single-particle Green's function is given by 
\begin{eqnarray}
G_c(k,\omega)=\frac{1}{\omega-\epsilon_c(k)-\Sigma^{\rm nor}(k,\omega)-W(k,\omega)},
\label{TCMSCGkw} 
\end{eqnarray}
with 
\begin{eqnarray}
W(k,\omega)=\frac{\Sigma^{\rm ano}(k,\omega)^2}{\omega+\epsilon_c(k)+\Sigma^{\rm nor}(k,-\omega)^*},
\label{TCMSCWkw} 
\end{eqnarray}
where
\begin{eqnarray}
\Sigma^{\rm nor}(k,\omega)=\frac{\Lambda(k)^2(\omega+\epsilon_d(k))}{\omega^2-\epsilon_d(k)^2-\Delta_d(k)^2},
\label{TCMSCSCSNkw} 
\end{eqnarray}
and 
\begin{eqnarray}
\Sigma^{\rm ano}(k,\omega)=\Delta_c(k)+\frac{\Lambda(k)^2\Delta_d(k)}{\omega^2-\epsilon_d(k)^2-\Delta_d(k)^2}.
\label{TCMSCSCSAkw} 
\end{eqnarray}

Now the pole position of $\Sigma^{\rm nor}$ at \txr{$\omega=\epsilon_d(k)$} in the normal state (expected to generate the pseudogap) (see Eq.(\ref{TCMNSkw})) is modified to  $\omega=\pm\sqrt{\epsilon_d(k)^2+\Delta_d(k)^2}$.
The residue of this pole in $\Sigma^{\rm nor}$ cancels with that of $W$ at the same $\omega$, which results in the cancellation of the poles in the normal and anomalous contributions in $G_c$~\cite{Sakai2016a}.

We employ the spectral function $A(k,\omega)$ given in the ARPES data reported by Kondo {\it et al.}~\cite{kondo2011disentangling}.
for Bi2212 and the successfully derived self-energies by machine learning~\cite{Yamaji2021}.
The obtained spectral function and the self-energy are fitted by the TCFM.
Hereafter, we take the energy unit by eV. In this unit, we obtain the parameters 
$t_{c1}=0.1953$, $t_{c2}=-0.0762$, $t_{d1}=-0.0100$, $t_{d2}=-0.0036$, $\mu_c=0.2875$, $\mu_d=0.0405$, $\Delta_{c0}=0$, $\Delta_{d0}=0.0570$, 
and 
$\Lambda(k)=\Lambda_0+\Lambda_1(\cos k_x+1)(\cos k_y+1)$
with $\Lambda_0=0.0658$
and  $\Lambda_1=-0.014$. 
These satisfy the Fermi surface position including $(\pi, 0.14\pi)$ as the antinodal (AN) point and $(0.37\pi, 0.37\pi)$ as the nodal (N) point in the ARPES measurement~\cite{kondo2011disentangling}. $\Lambda_0$ and $\Lambda_0+\Lambda_1$ essentially play the role of determining the hybridization at the AN and N points respectively.  
They also essentially reproduce the dispersion across the Fermi surface reported in Ref.~\citen{Chen2019}. 

\begin{figure}[h!]
  \begin{center}
    \includegraphics[width=7cm]{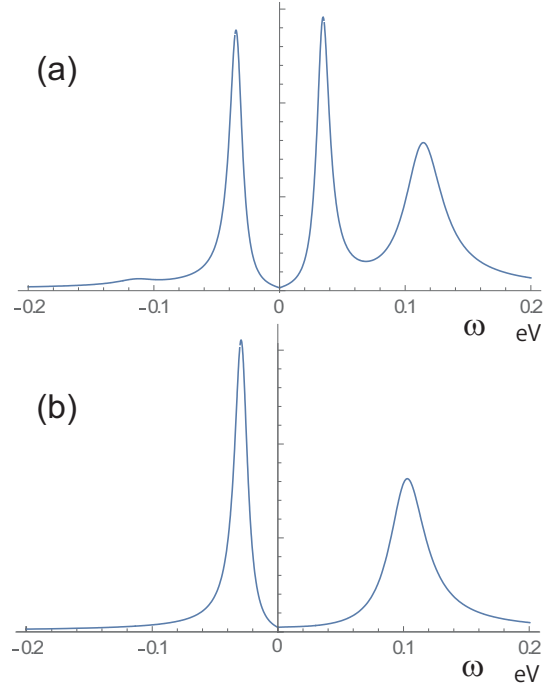}
    \end{center}
\caption{
(a) Spectral function $A(k,\omega)$ obtained from TCFM in the superconducting phase to reproduce the ARPES and the machine learning data. (b) $A(k,\omega)$ in the normal (pseudogap) state for the same parameters as (a) except for the choice  $\Delta_{d0}=\Delta_{c0}=0$.  
}
\label{Akw}
\end{figure}

\begin{figure}[h!]
  \begin{center}
    \includegraphics[width=7cm]{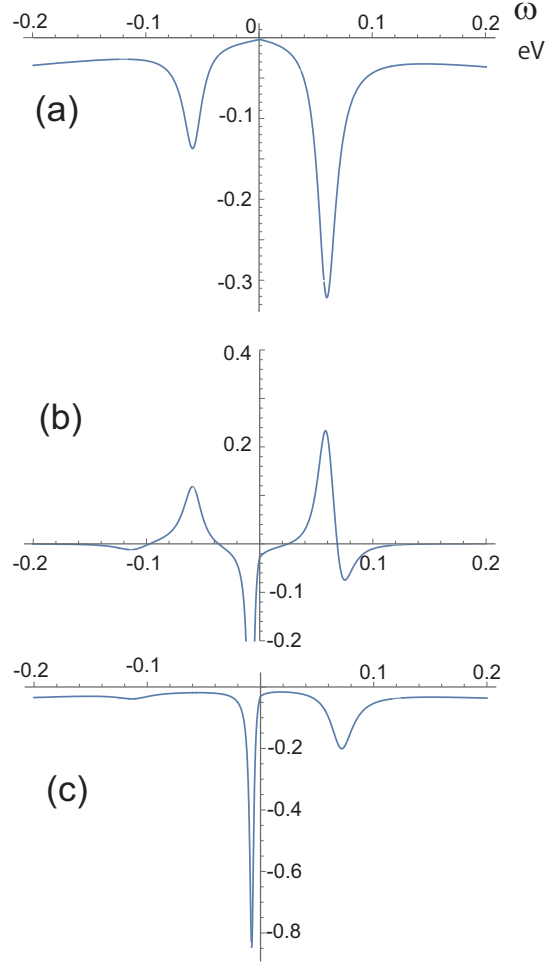}
    \end{center}
\caption{
(a) ${\rm Im} \Sigma(k,\omega)$ and (b) ${\rm Im} W(k,\omega)$ obtained from TCFM showing prominent either negative ((a)) or positive((b)) peaks at  $\omega\sim \pm 0.07$ eV.  
The peaks at $\sim \pm 0.07$ eV cancels in ${\rm Im} \Sigma(k,\omega)+{\rm Im} W(k,\omega)$ as is plotted in (c). 
}
\label{ImSigma}
\end{figure}

The broadening factor $\gamma(k,\omega)$ is taken to have the $k$-independent form 
\begin{eqnarray}
\gamma(\omega)=a|\omega|+b
\label{gamma}
\end{eqnarray}
with $a=0.15$ and $b=0.002$ to reproduce the broadening of both the peak and hump in the measured ARPES data analyzed below. This expression actually has the marginal Fermi liquid form~\cite{Varma}.
With these parameters, the spectral function $A(k,\omega)$ is illustrated 
in Fig.~\ref{Akw}. The peak, dip and hump positions and the superconducting gap amplitude at the AN Fermi surface point $(\pi, 0.14\pi)$ are seen as shown in Fig.~\ref{Akw}(a). 
Here, we can confirm that the peak, dip and hump energies, relative peak heights and peak widths in the ARPES data in the negative $\omega$ side for Bi2212 at the antinodal point well reproduce those measured in ARPES~\cite{kondo2011disentangling} and its machine learning result~\cite{Yamaji2021} by employing fitting values of the TCFM above. In the normal state shown in Fig.~\ref{Akw}(b), the gap structure identified as the pseudogap is essentially consistent with the ARPES data slightly above above $T_{\rm c}$~\cite{Kondo2015}. In the actual ARPES data, there exists presumable background effect as well as broad $\omega$ insensitive structure partly because of the extrinsic experimental conditions, which does not exist in the TCFM and we ignored it.
The obtained normal and anomalous self-energies of the obtained TCFM in the superconducting phase indeed have prominent peaks in the imaginary part, ${\rm Im}\Sigma^{\rm nor}$ and ${\rm Im} W$ at $\sim \pm 0.07$ eV as is seen in Fig.~\ref{ImSigma} and their peak cancels in the Green's function completely consistent with the machine learning result. 
Although an ambiguity exists, the fitting is better for $\Delta_{c0}=0$ rather than  $\Delta_{c0}$ dominant superconductivity to reproduce the self-energy structure estimated in the machine learning~\cite{Yamaji2021}, implying that the superconductivity is primarily triggered by the dark fermion contribution.
Note that the $d$-wave superconducting gap becomes nonzero even when $\Delta_{c0}=0$, because nonzero $\Delta_{d}$ generates the gap in the $c$ component as well through the momentum diagonal hybridization $\Lambda(k)$.
It should be noted that the pseudogap has an $s$-wave like full gap extending from the AN to N points as we see from the $k$ dependence of $\Lambda(k)$, consistently again with the previous model study~\cite{Sakai2016a}. 

%
\section{Formulation for RIXS spectra}
\begin{figure}[h!]
  \begin{center}
    \includegraphics[width=10cm]{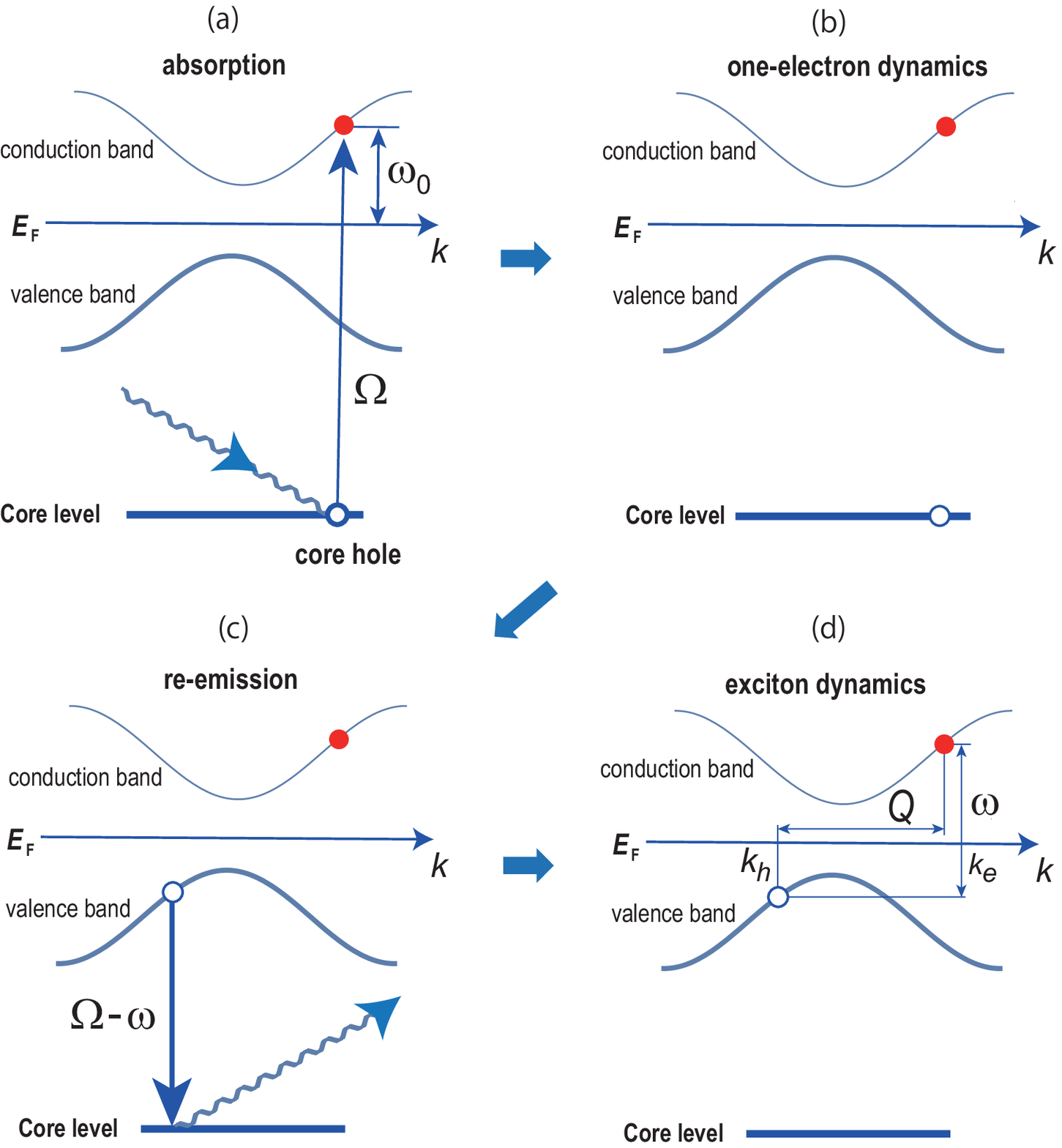}
    \end{center}
\caption{
Illustration of the electron dynamics in RIXS measurement. (a) Excitation of a core electron to the conduction band by X-ray absorption. (b)  Dynamics of the excited electron under the potential of the core hole. (c) Decay of a valence electron and recombination with the core hole accompanied by reemission of X-ray. (d) Dynamical evolution of electron-hole pair (exciton).
}
\label{RIXS_Illustration}
\end{figure}%

To proceed to studies by the integrated spectroscopy, we discuss now the consequence for the RIXS. The process of the RIXS dynamics is illustrated in Fig.~\ref{RIXS_Illustration}, where the X-ray excites a core electron to the conduction band and the subsequent decay of the valence electron to the core hole by X-ray emission makes an exciton (electron-hole pair), which evolves in time until the pair recombination. This RIXS dynamics intensity is expressed by
\begin{eqnarray}
I_{\rm RIXS}(Q,\omega, \omega_0; \sigma, \rho)&\propto& \sum_l|B_{li}(Q, \omega_0;\sigma,\rho)|^2\delta(E_l-\omega), \label{I_RIXS_1}
\end{eqnarray}
\begin{eqnarray}
B_{li}(Q,\omega_0;\sigma,\rho)=\sum_{m,j}e^{iQ\cdot R_m}\chi_{\rho,\sigma}\langle l|c_{m\sigma}|j\rangle\langle j|(\omega_0-E_j+i\Gamma)^{-1}|j\rangle\langle j |c^{\dagger}_{m\rho}|i(k=0)\rangle,
\label{I_RIXS_2} 
\end{eqnarray}
where $|i(k=0)\rangle$ is the ground state and $\chi_{\rho,\sigma}$ is the spin ($\rho$ and $\sigma$) dependent matrix representing the dipole transition amplitude.  The inverse of the core hole lifetime is denoted by $\Gamma$. Here, $\Omega$ is the energy of incident X-ray in Fig.~\ref{RIXS_Illustration}, but we subtract $E_{\rm F}-E_{\rm c}$ in the expression as $\omega_0 =\Omega -(E_{\rm F}-E_{\rm c})$ to measure energy from $E_{\rm F}$, where $E_{\rm F}$ is the Fermi energy and $E_{\rm c}$ is the core hole level. Here,  $c_m^{\dagger}$ is the local electron creation operator at the $m$-th site while $\omega_0=E_j$ is the excitation energy of an eigenstate $|j\rangle$ of the Hamiltonian $H_m=H+V_m$ for the system with one electron excited from the core to the conduction band at the $m$-th lattice site by the X-ray injection to the initial ground state. The excitation energy $E_j$ of the conduction electron is measured not from the core level but from the Fermi energy. The particle-hole (exciton) state $|l\rangle$ is characterized by the energy $E_l=\omega$ and the total momentum $Q$.  In the present TCFM, the created hole energy becomes $E_h=E_l-E_j$ (we take the hole creation energy positive).  Although we ignore the attractive potential of electron and hole in the exciton state in the TCFM, its effect can be approximately taken into account just by shifting $ \omega$ with the amount of the exciton binding energy $E_{\rm B}$ by interpreting as $ \omega \rightarrow  \omega - E_{\rm B}$. Here, the annihilation and creation of the core electron are not explicitly expressed in the equations and the core hole is treated just by the core hole potential $V_m$ at the $m$-th site in the intermediate state and we take $H$ as the TCFM (Eq.(\ref{TCfermionAnomalous})).  

Since the effect of nonzero $V_m$ is limited and qualitative feature does not change~\cite{Shi2017}, 
we take $V_m=0$.
In the translational invariant system with $V_m=0$,  Eq.(\ref{I_RIXS_2}) is rewritten as 
\begin{eqnarray}
B_{li}(Q,\omega_0;\sigma,\rho)&=&\sum_{j(k_e),k_h,k_e}\delta(Q-(k_h-k_e))\chi_{\rho,\sigma}\langle l(Q,k_e)|c_{k_h\sigma}|j(k_e)\rangle\langle j(k_e)|\nonumber \\
&&\times(\omega_0-E_j(k_e)+i\Gamma)^{-1} |j(k_e)\rangle\langle j(k_e) |c^{\dagger}_{k_e\rho}|i(k=0)\rangle,
\label{I_RIXS_3} 
\end{eqnarray}
where $|j(k_e)\rangle$ is the one-electron excited state with the specified momentum $k_e$ and $|l(Q,k_e)\rangle$ is the electron-hole (exciton) state with the electron momentum $k_e$ and the hole momentum $k_h=k_e+Q$.

In this paper, we consider spin non-flipping contribution $\Delta S=0$, which reduces to 
$B_{li}(Q,\omega;\sigma,\rho) \rightarrow B_{li}(Q,\omega; \uparrow, \uparrow)+B_{li}(Q,\omega; \downarrow, \downarrow)$. This can be measured in Cu $L_3$ or O $K$ edge in RIXS.
If $\Gamma$ can be taken as $\Gamma\rightarrow \infty$, $I_{\rm RIXS}$ reduces to the charge dynamical structure factor 
\begin{eqnarray}
N(Q,\omega)=\int dr_m dt e^{iQ\cdot r_m-i\omega t}\langle n_{0}(0)n_{m}(t)\rangle
\label{NQw} 
\end{eqnarray}
with the charge operator $n_{j}(t)=e^{-iH_mt}c_j^{\dagger}c_je^{iH_mt}$.
$\langle \cdots \rangle$ represents the ground state average.  

The single-particle electronic excitation structure in the unoccupied conduction states above the Fermi level can be probed by the X-ray absorption spectra (XAS) using the same X-ray source simultaneously as
\begin{eqnarray}
I_{\rm XAS}(\omega)=-\frac{1}{\pi}{\rm Im} \sum_{j,m,k}\langle 0|c_{m\sigma}(\omega-E_j(k)+i\Gamma_{\rm XAS}(k,\omega))^{-1}c^{\dagger}_{m\sigma}|i(k=0)\rangle,
\label{XAS} 
\end{eqnarray}
where  $\Gamma_{\rm XAS}(k,\omega)$ is a momentum-energy dependent damping constant.
%
%
\section{Results}
\subsection{XAS Spectra}
By using the parameters of the TCFM that reproduce the spectral function 
measured by ARPES as described above, we calculate the XAS and RIXS spectra 
both in the   normal (pseudogap) state with $\Delta_c=\Delta_d=0$ and the 
$d$-wave superconducting state and compare these two.
These will also be compared with the single-band case. 

For the fitted parameters, the XAS spectra are shown in Fig.~\ref{XAS} for the pseudogap phase (a) and for the superconducting phase (b).  This indicates that a large intensity of  quasiparticle states is found with two peak structures in the region $0<\omega <0.2$ eV and the broad intensity arising from the ingap states separated by the $s$-wave like pseudogap from the quasiparticle band is found in $0.2<\omega <1.5$ eV. Here, the quasiparticle band and the ingap states are the consequence of the diagonalization of the TCFM and the pseudogap corresponds to the hybridization gap generated by $\Lambda(k)$. Because the TCFM ignores the upper Hubbard band, we can interpret that the high energy part near 1.5 eV partially represents the upper Hubbard weight in a compromised fashion. In the realistic XAS measurements, the damping $\Gamma_{\rm XAS}(k,\omega)$ is essentially the same as $\Gamma$ in Eqs.(\ref{I_RIXS_2}) and (\ref{I_RIXS_3}), which depends on the natural width of the core electron level as is discussed below and may give broader peaks than those in Fig.~\ref{XAS}. Here, we intend to figure out the intrinsic feature of the excitation spectra of conduction bands by taking an artificially small $\Gamma_{\rm XAS}$.
\begin{figure}[h!]
  \begin{center}
    \includegraphics[width=7cm]{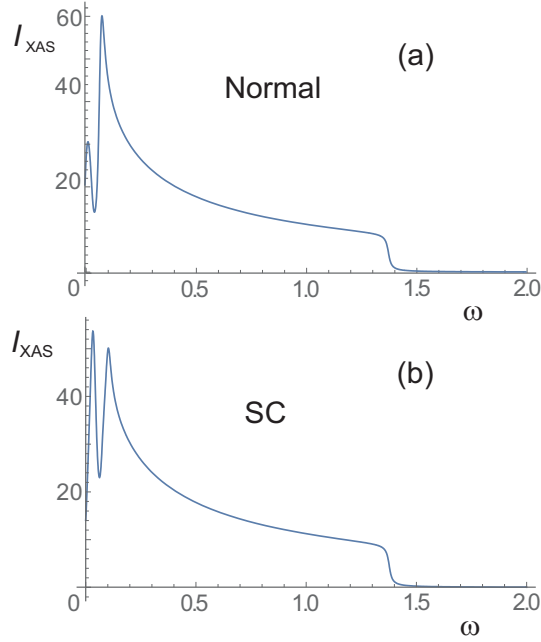}
    \end{center}
\caption{
Energy dependence of XAS spectra  in the normal (pseudogap) phase (upper panel, (a)) and $d$-wave superconducting phase with $\Delta_{d0}=0.057$eV and $\Delta_{c0}$=0 (lower panel, (b)).  For simplicity, we take $\Gamma_{\rm XAS}(k,\omega)=0.01$eV to be $k$ and $\omega$ independent form. 
}
 \label{XAS}
\end{figure}%

\subsection{RIXS Spectra in the Limit of Charge Structure Factor}
We next study the RIXS spectra and consider the case where an electron is excited from the core to the energy above the Fermi level $E_{\rm F}$ (namely $E_j>0$) and a hole is created below $E_{\rm F}$ (accompanying the decay of an electron to the core hole.) Note that we set $E_{\rm F}=0$.
In the superconducting state, other cases can contribute in principle to the RIXS intensity through the anomalous component of the Bogoliubov quasiparticle, but its effect must be limited only for small $\Delta \omega$ less than or comparable to the superconducting gap in actual experiments and we ignore it.   
In the following, we introduce a small smearing factor $\eta=0.005$ eV instead of $\delta$-function in Eq.(\ref{I_RIXS_1}).

\begin{figure}[h!]
  \begin{center}
    \includegraphics[width=7cm]{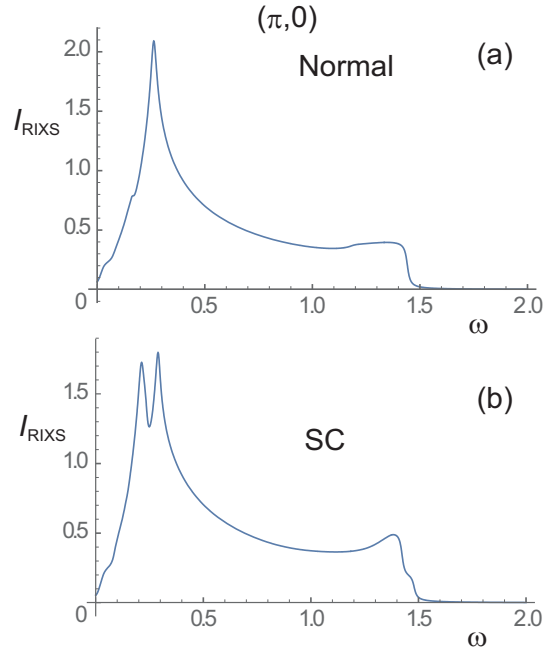}
    \end{center}
\caption{
Energy dependence of RIXS spectra $I_{\rm RIXS}$ at $Q=(\pi,0)$ for $\Gamma=3.0$ eV in (a) normal (pseudogap) phase (upper panel), (b) $d$-wave superconducting phase
for $\Delta_{d0}=0.057$ eV and $\Delta_{c0}$=0.   The incident energy $\omega_0$ is 0.076 eV for the pseudogap case and 0.032 eV for the superconducting case.
}
\label{Gamma3_Pi0_N_SC}
\end{figure}

Let us discuss the RIXS spectra first for large $\Gamma$, in the region essentially the same as the limit $\Gamma\rightarrow\infty$, where the RIXS intensity  is reduced to the charge structure factor (Eq.(\ref{NQw})).
More specifically we take $\Gamma=3.0$ eV and show results in the case of initial excitation of an electron to the conduction band with $E_j$ in Eqs. (\ref{I_RIXS_1}) and (\ref{I_RIXS_2}) at the peak position of XAS spectra. Namely, $E_j=0.076$ eV for the pseudogap phase and $E_j=0.032$ eV for the superconducting phase, because experimentally the excitation to the energy having the largest density of states makes the signal detection easier. However, the RIXS intensity does not sensitively depend on $E_j$, if it is small $<0.2$ eV. We will examine the incident-energy dependence later. Figure~\ref{Gamma3_Pi0_N_SC} shows the RIXS intensity for the momentum transfer $Q=(\pi,0)$, which has a prominent peak around $\omega\sim 0.26$ eV followed by a long tail extended to $\sim 1.5 $ eV with a shoulder at $1.4$ eV in the pseudogap state shown in (a). In the superconducting phase in (b), the peak at 0.26 eV splits into two peaks ascribed to the superconducting gap formation.  This splitting scales with the superconducting gap amplitude of the dark fermion $\Delta_{d0}$ at the AN point. In addition, the shoulder at $\sim 1.4$ eV becomes more prominent and develops a small peak in the superconducting state. The shoulder around 1.4 eV mainly comes from the excitation of electron by the X-ray to the region $(\pi,\pi)$ followed by the creation of a hole around $(0,\pi)$, which results in the exciton momentum $Q=(\pi,0)$.  

Even when the dark fermion is absent (namely by switching off the hybridization $\Lambda(k)$), the RIXS intensity is qualitatively similar as that seen in Fig.~\ref{Gamma3_Pi0_N_SC}(a) in the normal state (not shown), which means that the pseudogap does not have an appreciable effect in the normal state, if the incident energy is small and $\Gamma$ is large as in this case. The same applies to other $Q$. We will discuss this issue later when we discuss the case of small $\Gamma$, which shows qualitative difference from the single-band case.

\begin{figure}[h!]
  \begin{center}
    \includegraphics[width=7cm]{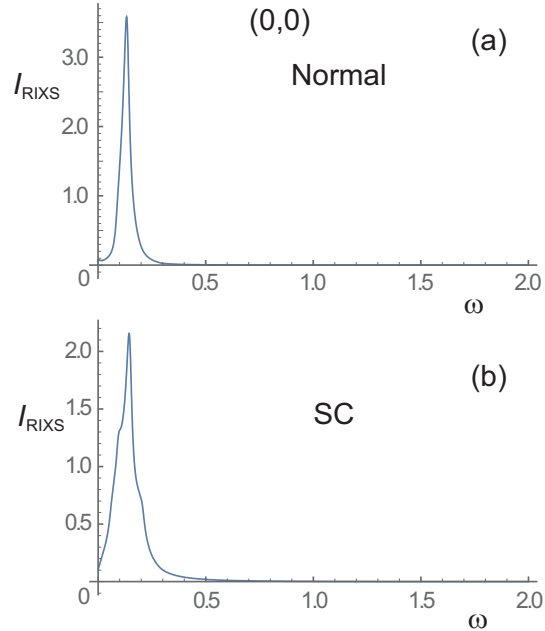}
    \end{center}
\caption{
Energy dependence of RIXS spectra $I_{\rm RIXS}$ at $Q=(0,0)$ for $\Gamma=3.0$ eV in (a) normal (pseudogap) phase (upper panel), (b) $d$-wave superconducting phase
for $\Delta_{d0}=0.057$ eV and $\Delta_{c0}$=0.  The incident energies are the same as Fig.~\ref{Gamma3_Pi0_N_SC}.  
}
\label{Gamma3_00_N_SC}
\end{figure}

For the momentum transfer $Q=(0,0)$, the spectra is shown in Fig.~\ref{Gamma3_00_N_SC}. The intensity is concentrated as a low-energy peak arising from the transition between the quasiparticle and ingap states representing the direct gap corresponding to the infrared structure in the optical conductivity.
The peak energy corresponds to the averaged direct $s$-wave-like psudogap energy scale.

\begin{figure}[h!]
  \begin{center}
    \includegraphics[width=7cm]{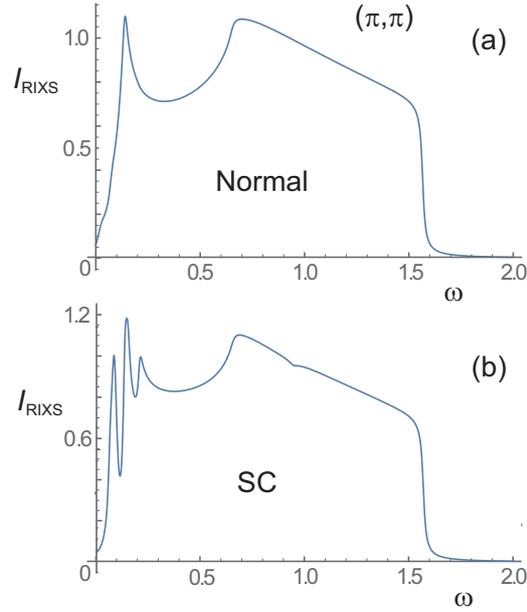}
    \end{center}
\caption{
Energy dependence of RIXS spectra $I_{\rm RIXS}$ at $Q=(\pi,\pi)$ for $\Gamma=3.0$ eV  in (a) normal (pseudogap) phase (upper panel), (b) $d$-wave superconducting phase
for $\Delta_{d0}=0.057$ eV and $\Delta_{c0}$=0.   The incident energies are the same as Fig.~\ref{Gamma3_Pi0_N_SC}.  
}
\label{Gamma3_PiPi_N_SC}
\end{figure}

The case of the momentum transfer $Q=(\pi,\pi)$ is shown in Fig.~\ref{Gamma3_PiPi_N_SC}, where the tail found for $Q=(\pi,0)$ is replaced by a broad prominent plateau-like structure in the region $0.5 <\omega <1.5$ eV and a more complicated splitting of the peaks than the case $Q=(\pi,0)$ around 0.14 eV is seen in the superconducting phase.  The plateau contains all the contribution including the transitions from $(0,0)$ to $(\pi,\pi)$ and the peak splitting reflects the transitions between the states including gapped AN points namely between the $(0,\pi)$ and $(\pi,0)$ regions as well as the transitions between the N points around $(\pm \pi/2,\pm \pi/2)$. 
Inspections of intermediate $Q$ points in the Brillouin zone show continuous evolution of the typical spectral features shown here from $(0,0)$, and $(\pi,0)$ through $(\pi,\pi)$.

\subsection{Beyond Charge Structure Factor}
Since experimental $\Gamma$ does not satisfy the large $\Gamma$ condition, we next study the case of small $\Gamma$ to make the prediction for the experiment more relevant. Actually, RIXS spectra at small $\Gamma$ are expected to contain richer information of the exciton dynamics, because the momenta of electron and hole can separately be resolved and extracted. See also Ref.\citen{Tohyama2018} for theoretical studies on the $\Gamma$ dependence of RIXS spectra, where the more prominent peak at large energy for the larger $\Gamma$ is similar. Below we study the case $\Delta_{d0}=0.05$ instead of $\Delta_{d0}=0.057$ for simplicity to understand the qualitative trend and see the dependence on the energy of the excited conduction electron $E_j$ regardless of the XAS peak energy.

\begin{figure}[h!]
  \begin{center}
    \includegraphics[width=7cm]{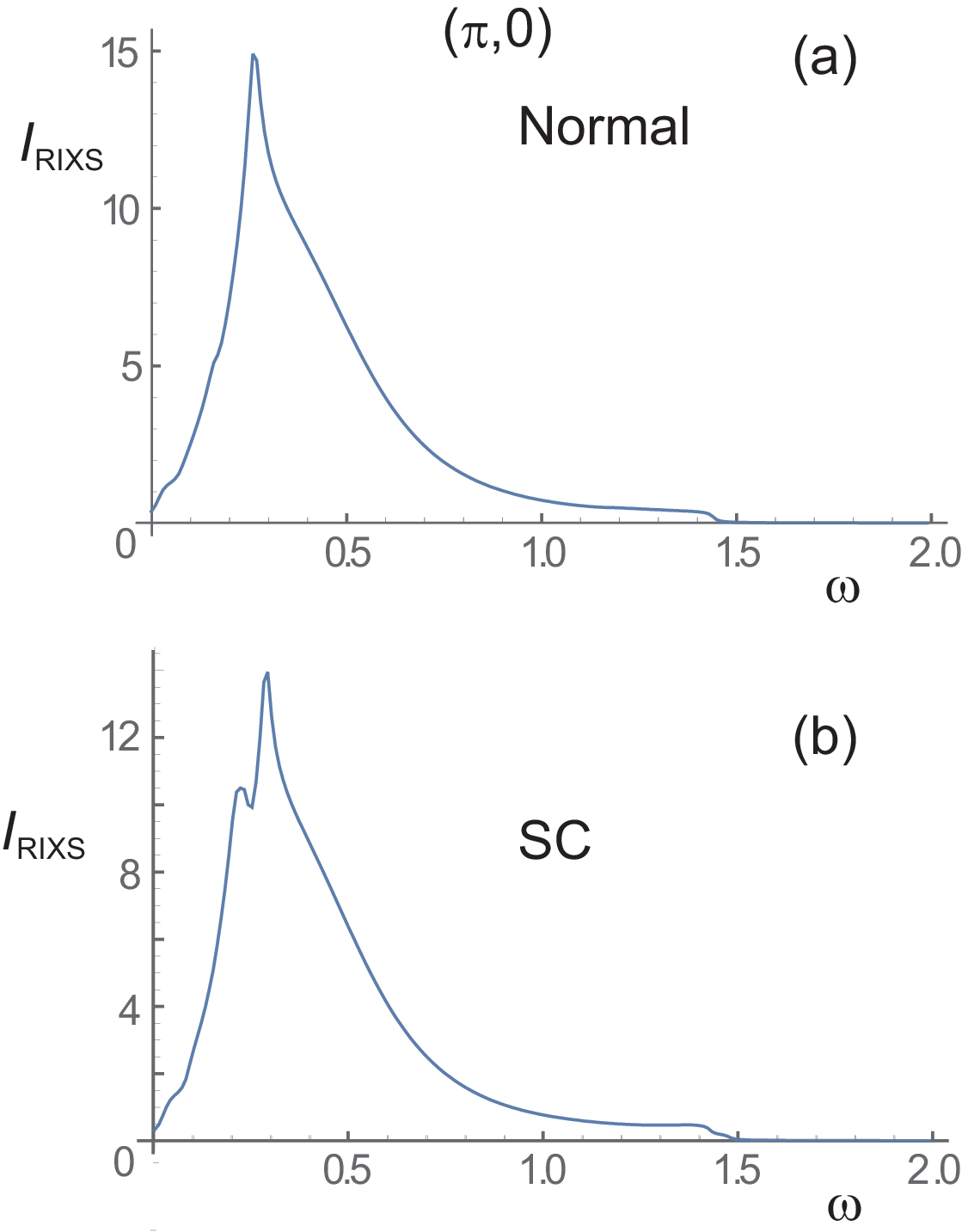}
     \includegraphics[width=7cm]{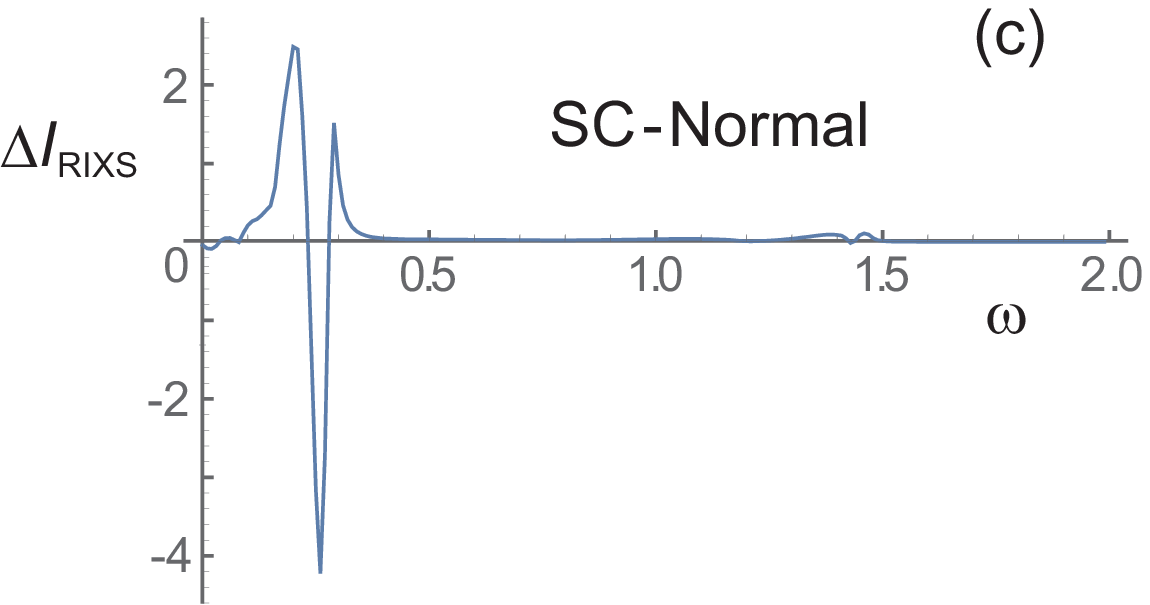}
   \end{center}
\caption{
Energy dependence of RIXS spectra $I_{\rm RIXS}$ at $Q=(\pi,0)$ for $\Gamma=0.3$ eV and the incident energy $\omega_0=0.3$ eV in (a) normal (pseudogap) phase (upper panel), (b) $d$-wave superconducting phase and (c) their difference $\Delta I_{\rm RIXS}$ for $\Delta_{d0}=0.05$ eV and $\Delta_{c0}$=0.  
}
\label{Gamma0.3_0.3Pi0_N_SC}
\end{figure}

Since $\Gamma$ for the RIXS measurement using Cu $L_3$ edge is estimated to be around 0.28 eV~\cite{Krause1979}, we next discuss the case $\Gamma=0.3$ eV. 
Figure~\ref{Gamma0.3_0.3Pi0_N_SC} shows the RIXS intensity at $Q=(\pi,0)$ for the incident energy 0.3 eV. 
It has a peak at low energies $\sim 0.3$ eV similarly to the case of $\Gamma=3.0$ eV.  and the increment $\Delta I_{\rm RIXS}=I_{\rm RIXS}^{\rm SC}-I_{\rm RIXS}^{\rm PG}$ from the pseudogap to the superconducting phase is appreciable basically only around 0.3 eV, where the peak splitting by the superconducting gap formation makes the positive-negative-positive peak structure in $\Delta I_{\rm RIXS}$ with total weight nearly unchanged between $I_{\rm RIXS}^{\rm SC}$ and $I_{\rm RIXS}^{\rm PG}$.

\begin{figure}[h!]
  \begin{center}
    \includegraphics[width=7cm]{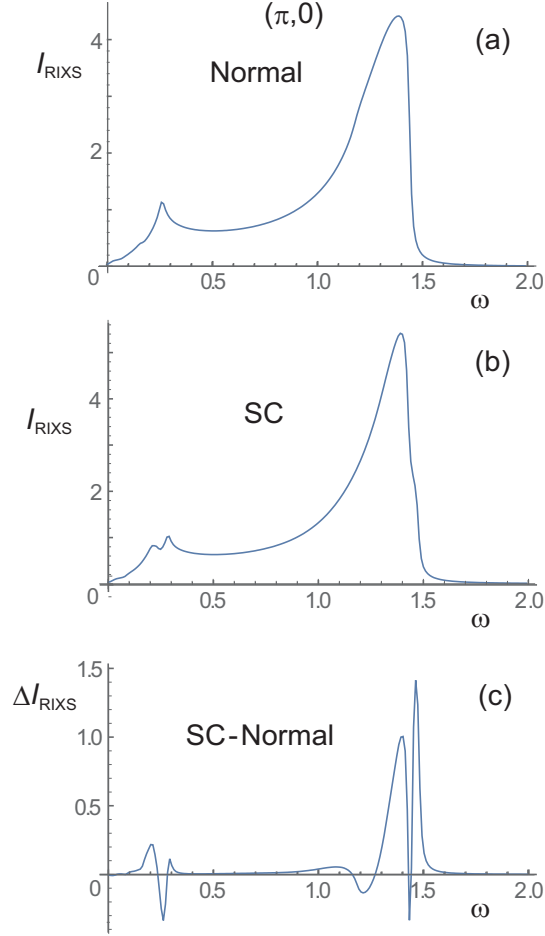}
    \end{center}
\caption{
Energy dependence of RIXS spectra $I_{\rm RIXS}$ at $Q=(\pi,0)$ for $\Gamma=0.3$ eV and the incident energy $\omega_0=1.37$ eV in (a) normal (pseudogap) phase (upper panel), (b) $d$-wave superconducting phase and (c) their difference $\Delta I_{\rm RIXS}$ for $\Delta_{d0}=0.05$ eV and $\Delta_{c0}$=0.  
}
\label{Gamma0.3_1.37Pi0_N_SC}
\end{figure}

\begin{figure}[h!]
  \begin{center}
    \includegraphics[width=7cm]{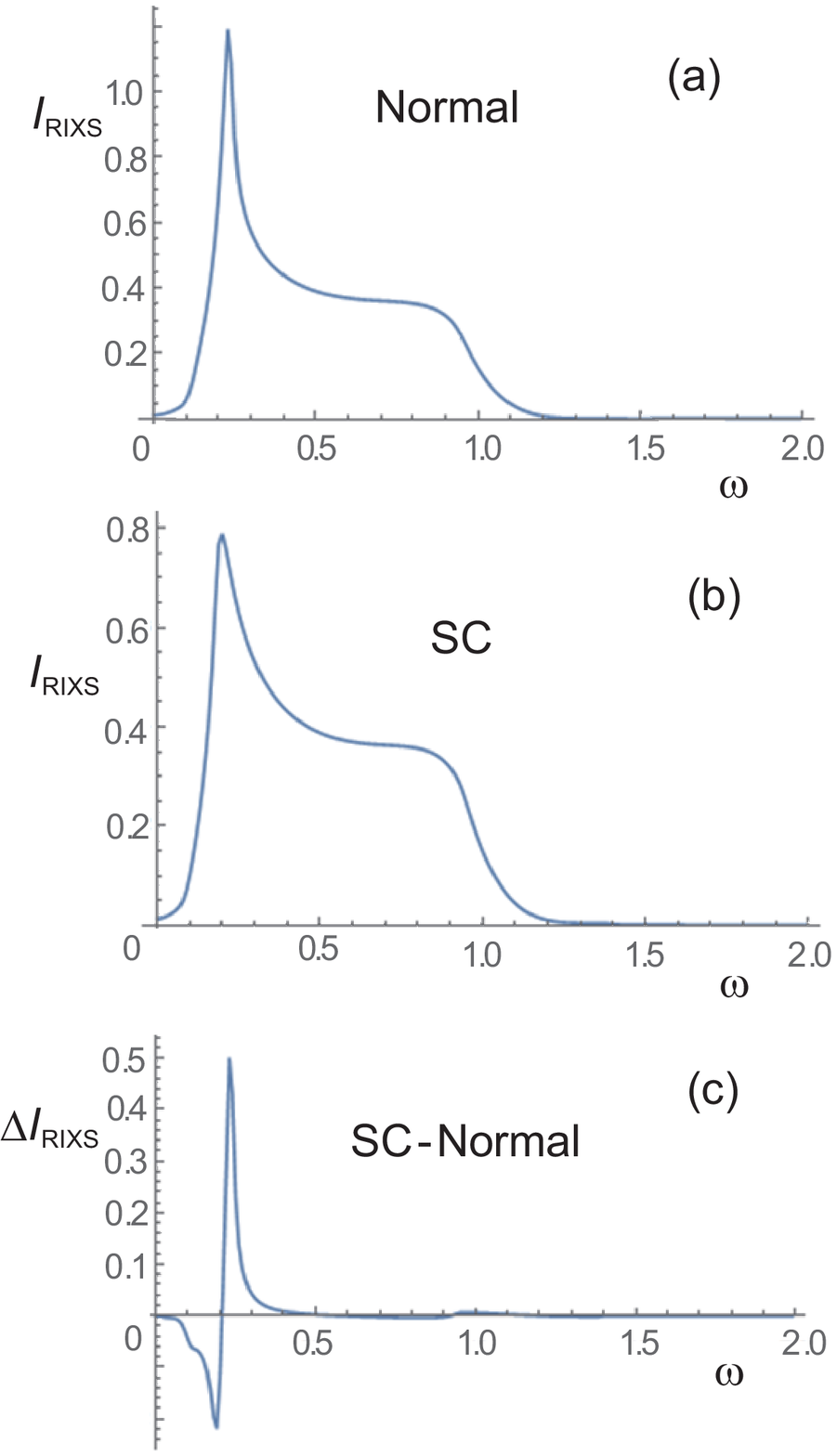}
    \end{center}
\caption{
Energy dependence of RIXS spectra $I_{\rm RIXS}$ at $Q=(\pi,0)$ for $\Gamma=0.3$ eV and the incident energy $\omega_0=1.37$ eV for the single-band case, where $\Lambda(k)$ is switched off to 0.  (a) normal (pseudogap) phase (upper panel), (b) $d$-wave superconducting phase and (c) their difference $\Delta I_{\rm RIXS}$ for $\Delta_{d0}=0$ and $\Delta_{c0}=0.05$.  
}
\label{Gamma_single-band_0.3_1.37Pi0_N_SC}
\end{figure}

However, if the incident energy increases and the electron is excited around the top of the ingap state ($\sim 1.37$ eV) by X-ray in the present TCFM, the RIXS intensity shows a sharp peak around the incident energy and it is clearly seen that the RIXS weight is enhanced in total in the superconducting state relative to the normal (pseudogap) state at around the incident energy as one can see in Fig.~\ref{Gamma0.3_1.37Pi0_N_SC}.
The enhancement by the superconductivity with the gap size 0.05 eV at such a high energy $>1$ eV is a remarkable and nontrivial property of this fractionalized two-component model. For the single-band case, such an enhancement does not occur
as one sees in Fig.~\ref{Gamma_single-band_0.3_1.37Pi0_N_SC}. 

\begin{figure}[h!]
  \begin{center}
    \includegraphics[width=7cm]{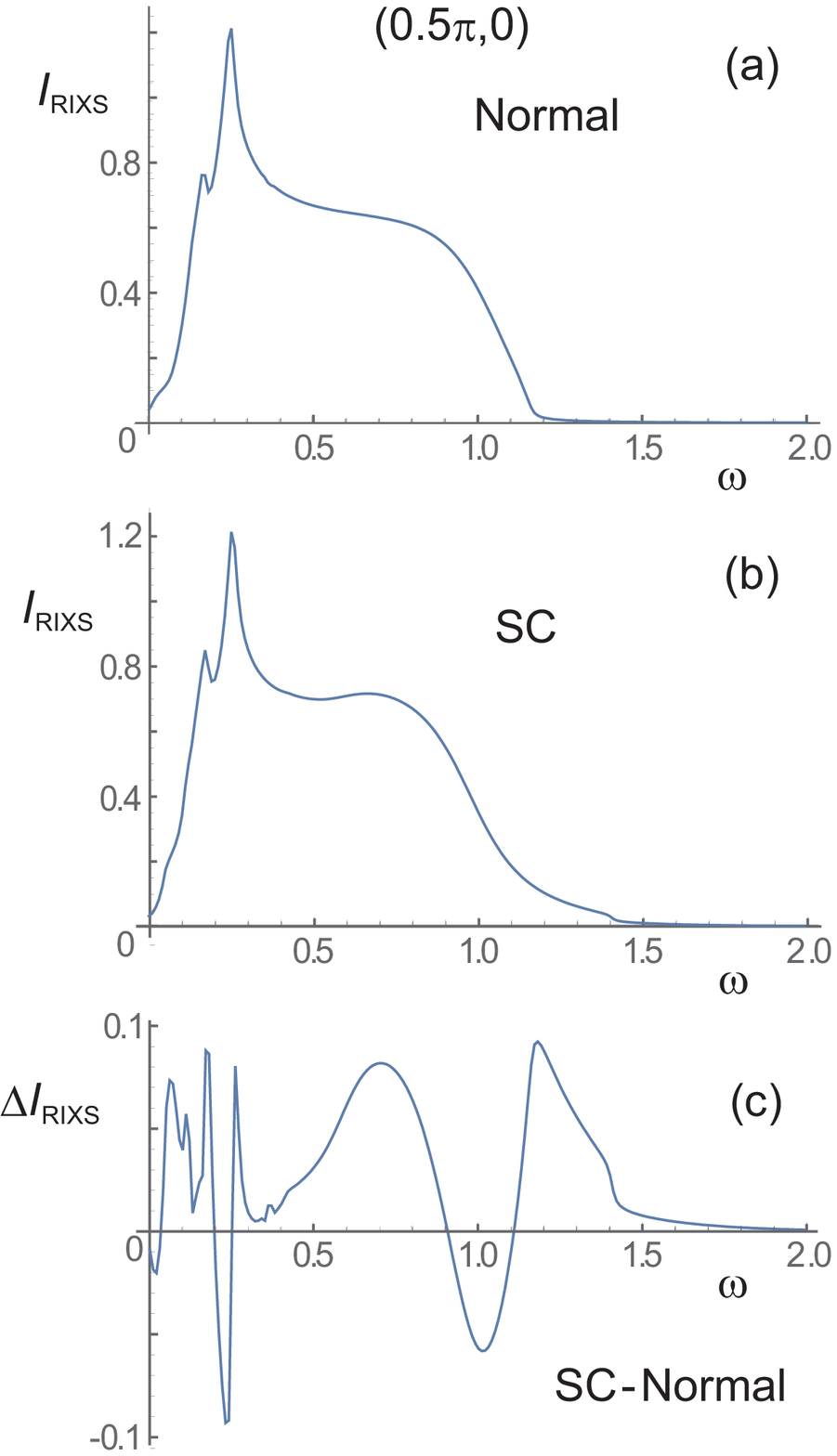}
   \end{center}
\caption{
Energy dependence of RIXS spectra $I_{\rm RIXS}$ at $Q=(0.5\pi,0)$ for $\Gamma=0.3$ eV and the incident energy $\omega_0=1.37$ eV in (a) normal (pseudogap) phase (upper panel), (b) $d$-wave superconducting phase and (c) their difference $\Delta I_{\rm RIXS}$ for $\Delta_{d0}=0.05$eV and $\Delta_{c0}$=0.  
}
\label{Gamma0.3_1.37_0.5Pi_0_N_SC}
\end{figure}

\begin{figure}[h!]
  \begin{center}
    \includegraphics[width=7cm]{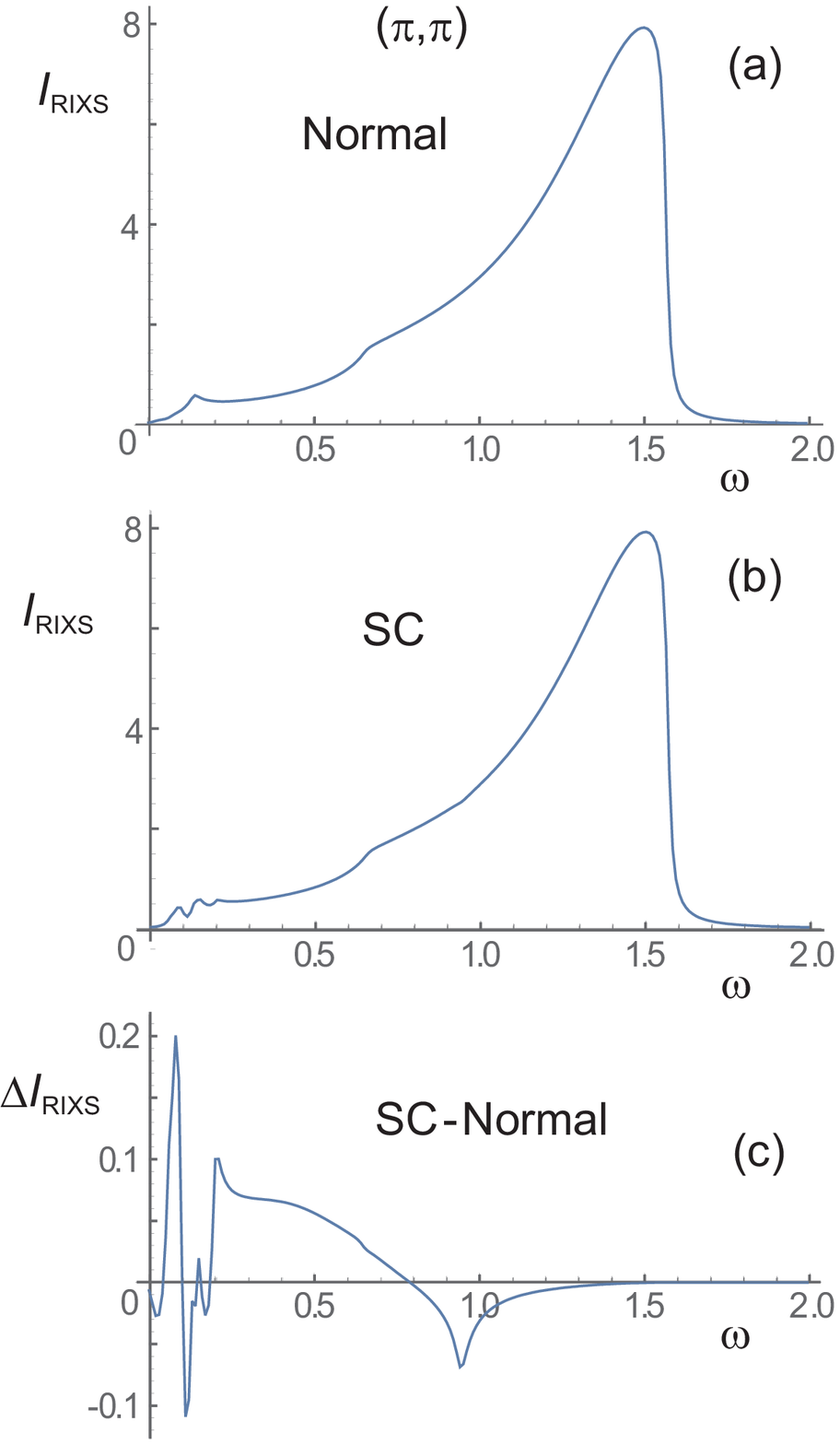}
   \end{center}
\caption{
Energy dependence of RIXS spectra $I_{\rm RIXS}$ at $Q=(\pi,\pi)$ for $\Gamma=0.3$ eV and the incident energy $\omega_0=1.37$ eV in (a) normal (pseudogap) phase (upper panel), (b) $d$-wave superconducting phase and (c) their difference $\Delta I_{\rm RIXS}$ for $\Delta_{d0}=0.05$eV and $\Delta_{c0}$=0.  
}
\label{IRIXS_Gamma0.3_1.37_Pi_Pi_N_SC}
\end{figure}

This enhancement disappears for small $Q$, but even at $(0.5\pi,0)$, a small enhancement is still seen as in Fig.~\ref{Gamma0.3_1.37_0.5Pi_0_N_SC}. This is helpful if the available $Q$ is limited in the experiments.
When $Q$ is in the diagonal direction, around $(\pi,\pi)$, though the peak is prominent around the incident energy as in Fig~\ref{IRIXS_Gamma0.3_1.37_Pi_Pi_N_SC}, the enhancement becomes less clear, while the enhancement is tiny but visible for smaller $Q$ such as $Q=(0.5\pi,0.5\pi)$(not shown).  

\begin{figure}[h!]
  \begin{center}
    \includegraphics[width=7cm]{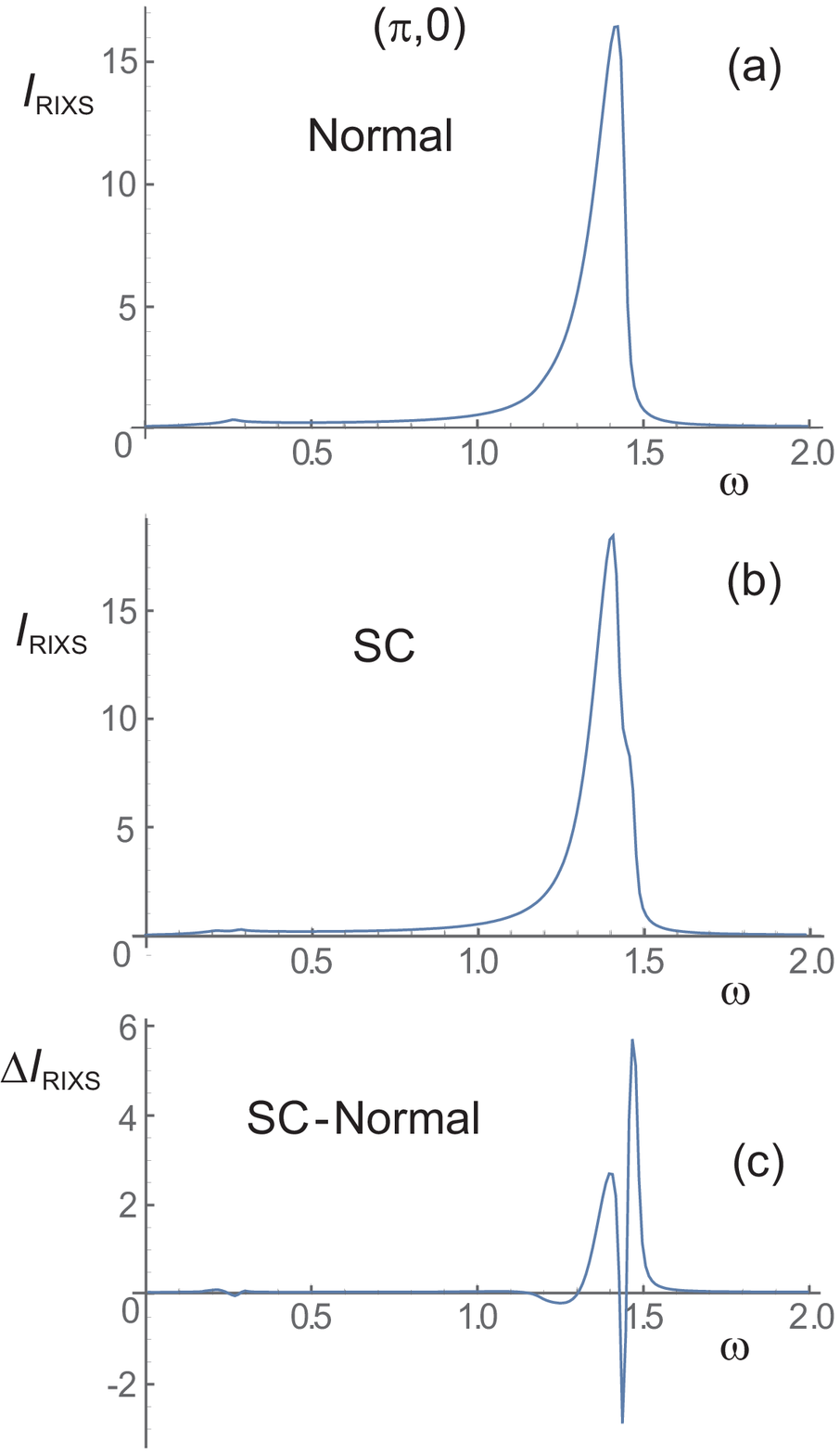}
   \end{center}
\caption{
Energy dependence of RIXS spectra $I_{\rm RIXS}$ at $Q=(\pi,\pi)$ for $\Gamma=0.07$ eV and the incident energy $\omega_0=1.37$ eV in (a) normal (pseudogap) phase (upper panel), (b) $d$-wave superconducting phase and (c) their difference $\Delta I_{\rm RIXS}$ for $\Delta_{d0}=0.05$eV and $\Delta_{c0}$=0.  
}
\label{IRIXS_Gamma0.07_1.37_Pi_0_N_SC}
\end{figure}

This enhancement becomes relatively even more prominent in the superconducting phase, when $\Gamma$ is further decreased:
When we reduce $\Gamma$ to 0.07 eV, which is similar to the value $\sim 0.072$ eV for oxygen $K$ edge absorption for $1s\rightarrow 2p$ transition~\cite{Menzel},
Fig.~\ref{IRIXS_Gamma0.07_1.37_Pi_0_N_SC} shows more prominent enhancement by the superconductivity at the incident energy 1.37 eV than the case of larger $\Gamma$, in the ratio to the peak intensity in the pseudogap state at $Q=(\pi,0)$. The reason for the larger enhancement in small $\Gamma$ is that the combination of $(\pi,0)$ hole and $(\pi,\pi)$ electron is selectively picked up for $Q=(\pi,0)$ without broadening, where the effects of the fractionalization is larger as we discuss in the next section. This selective enhancement is broadened and smeared if $\Gamma$ gets large and a small enhancement can be seen even when $\omega_0$ is at the peak of XAS. To detect the enhancement easily, however, smaller $\Gamma$ with the tuned $\omega_0$ at the energy of the ingap top at $(\pi,\pi)$ is better. 

%
%
\section{Discussion}
\begin{figure}[h!]
  \begin{center}
    \includegraphics[width=7cm]{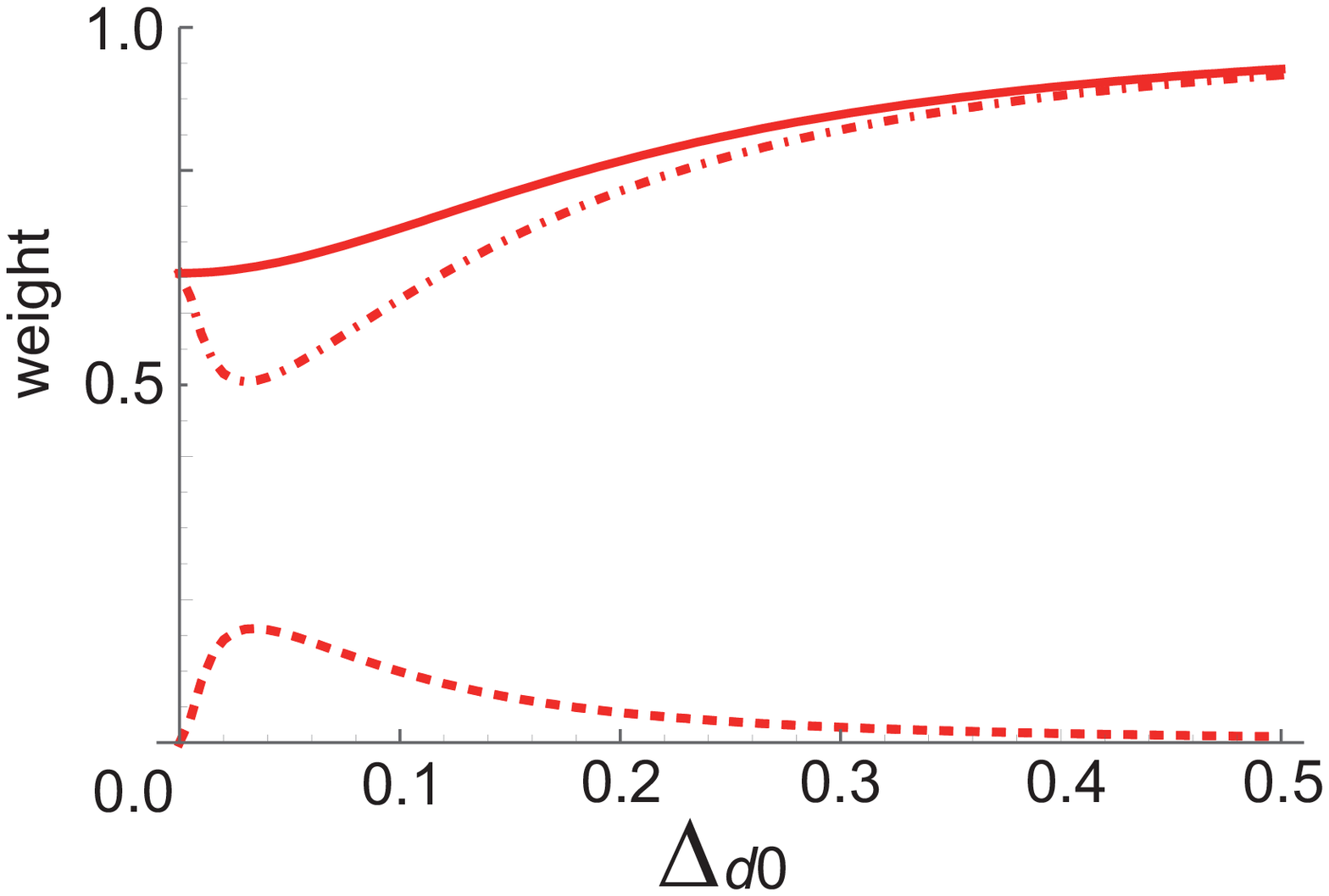}
   \end{center}
\caption{
Weight of $c$ component in the wave functions with negative energies in TCFM at the AN point. Dashed and dash-dotted curves are the weight of lower and higher energy states at the energy $<0$, respectively. Solid curve is the total weight. 
}
\label{WFcomponent_weightAN_c}
\end{figure}

The mechanism of the enhancement, we propose, is the following: For the excited particle at $\omega_0=E_j$, if $E_j$ corresponds to the region near the top of the ingap dispersion, (note that in the present case $E_j= 1.37$ eV corresponds to the ingap band top at $(\pi,\pi)$), the dominant contribution is always the $c$ component, which does not change between the pseudogap and superconducting state as one easily expects. On the other hand, the dominant contribution for the hole in the partner of the excited exciton is found around the AN point including $(\pm\pi, 0)$ and $(0, \pm\pi)$. The two bands below the Fermi level contributing to the hole creation around the AN point are dominantly $d$-hole like in one band and $c$-electron like in the other in terms of TCFM. 
Only the $c$-electron like state below the Fermi level contributes to annihilate an electron to decay to the core hole. With increasing the pairing for $d$ fermion, the total weight of the $c$ component for the wave functions with the energy below the Fermi level increases toward 1.0 as one sees in Fig.~\ref{WFcomponent_weightAN_c}, which causes the increase in the RIXS weight at $(\pi,0)$ in the superconducting state for the incident energy at around the top of the ingap state at $(\pi,\pi)$. 
 On the contrary, the weight of the particle component for $c$ fermion in the single-component systems (for instance in the case $\Lambda(k)=0$), does not change in total and the RIXS weight stays the same between the normal (pseudogap) and superconducting states.  The absence of appreciable enhancement in the superconducting phase for $(\pi,\pi)$ even at large $\omega$ (Fig.~\ref{IRIXS_Gamma0.3_1.37_Pi_Pi_N_SC}) is also consistent with the fact that the peak at large $\omega$ is mainly contributed from the transition between $(0,0)$ and $(\pi,\pi)$, where the states at these momenta mainly consist of the $c$ component and does not change between the normal (pseudogap) and the superconducting states. 
Therefore, when the band top of the conduction electron (either the ingap or the upper Hubbard) is excited in the superconducting state for the above choice of $Q$ and $\omega_0$, the enhancement of the RIXS weight is a clear indication of the electron fractionalization representable by the TCFM. 

The fractionalization implies that substantial degrees of freedom of electrons are hidden in the experimental measurements, which is detected as the incoherent part of electrons in transport and optical measurements. The physical entity of the fractionalized part, namely the dark fermion, is an open  issue to be clarified in the future, but the enhancement if observed indicates that the hidden (dark) particle or hole does not directly make an exciton with the visible electron or hole degrees of freedom. It also indicates that the hidden part has hybridization with the normal electron (or quasiparticle) and it also dominantly develops the superconducting gap over the conventional electron degrees of freedom.  In this understanding, the real visible superconducting gap in the $c$ part is drugged and driven by the gap formation of the dark fermion, $d$. 

Bi2212 is a bilayer system, and in principle it has a bonding and a antibonding band. One might think that these two components may equally show the enhancement. However, in this case, the superconducting transition is driven by the band near the fermi level and the increase in the weight or restructuring of this band does not occur. So, the enhancement if it occurs can also exclude the contribution of this trivial  two-component system from the contribution of the present fractionalization.   
%
%
\section{Summary and Outlook}
We have studied the RIXS intensity and have shown that the intensity is enhanced in the superconducting phase from the normal (pseudogap) phase for the appropriate choices of the incident energy $\omega_0$ and the momentum transfer $Q$, if the electron fractionalization captures the essence of low-energy charge excitation based on the two-component fermion description.  The parameters for the two-component description are derived and fitted from the ARPES measurement of Bi2212 cuprate superconductor combined with the machine learning study to infer the structure of the normal and anomalous self-energies of the single-particle Green's function. The predicted RIXS spectra can be measured and critically tested in the experiment with the present improved accuracy for the O $K$ and Cu $L_3$ edges, by picking up the spin non-flip charge excitations to examine the validity of the present type of fractionalization. Since the single-band description of the superconducting phase does not show such enhancement, the presence or absence of the  enhancement can indeed be used as a stringent test for the electron fractionalization in the cuprate superconductors. 

If the line shape of the RIXS intensity is clarified with high resolution, the self-energy of the exciton can be extracted beyond the TCFM. This will serve to clarify the entity of the dark fermion as well as the dynamics of the exciton, which is useful to directly extract the superconducting mechanism. 

Combined analyses of independent spectroscopic measurements with the help of the theoretical and computational analyses including the newly developed machine learning attempted in the present study will provide us in more general perspective with a powerful way to overcome the limitation and uncertainty we are facing with in conventional analyses based on single and independent probe measurement for strongly correlated electron systems for decades.  The approach exemplified here  may open the new direction of research as ``{\it integrated spectroscopy}". 

{\bf Acknowledgements}
The author thanks useful discussions with Atsushi Fujimori and Di-Jing Huang.
This work was financially supported by Grant-in-Aids
for Scientific Research (JSPS KAKENHI) (No. 16H06345) from Ministry of
Education, Culture, Sports, Science and Technology (MEXT), Japan. This work
was also supported in part by the projects conducted
under MEXT Japan named as ``Priority Issue
on Post-K computer" and ``Program for Promoting Research
on the Supercomputer Fugaku" in the subproject, ``Basic Science
for Emergence and Functionality in Quantum Matter:
Innovative Strongly-Correlated Electron Science by
Integration of Fugaku and Frontier Experiments".  We
also thank the support by the RIKEN Advanced Institute for Computational
Science through the HPCI System Research project (hp190145, and hp200132)
supported by MEXT.

\bibliographystyle{jpsj_mod}
\bibliography{Resonant_Inelastic_Xray_Spectra_TCFM.0510_arXiv}
\end{document}